\documentclass[aps,twocolumn,showpacs]{revtex4}
\usepackage{epsfig}
\usepackage{graphicx}
\usepackage{amsmath}
\usepackage{mathrsfs}
\usepackage{dcolumn}
\newcommand{\be}{\begin{equation}}
\newcommand{\ee}{\end{equation}}
\newcommand{\bey}{\begin{eqnarray}}
\newcommand{\eey}{\end{eqnarray}}
\newcommand{\ba}{\begin{array}}
\newcommand{\ea}{\end{array}}
\newcommand{\bi}{\begin{itemize}}
\newcommand{\ei}{\end{itemize}}
\newcommand{\bem}{\begin{enumerate}}
\newcommand{\eem}{\end{enumerate}}
\newcommand{\bw}{\begin{widetext}}
\newcommand{\ew}{\end{widetext}}
\newcommand{\ra}{\rangle}
\newcommand{\la}{\langle}

\newcommand{\ww}{\widetilde}

\newcommand{\bR}{{\bf R}}
\newcommand{\bx}{{\bf x}}

\begin{document}

 \title{
 Stability of Fock states in a two-component Bose-Einstein condensate with a regular
 classical counterpart}

\author{Wen-ge Wang$^{1,2}$, Jie Liu$^3$, and Baowen Li$^{2,4}$}
\affiliation{
 $^1$ Department of Modern Physics, University of Science and Technology of China,
 Hefei 230026, China
 \\ $^2$ Department of Physics and Centre for Computational Science and Engineering,
 National University of Singapore, 117542, Singapore
 \\ $^3$ Institute of Applied Physics and Computational Mathematics, P.O.Box
 100088, Beijing, China
 \\  $^4$Graduate School for Integrative Sciences and Engineering, National University of Singapore,
 117597, Singapore
 }

 \date{\today}

 \begin{abstract}

 We study the stability of a two-component Bose-Einstein condensate (BEC)
 in the parameter regime in which its classical counterpart has regular motion.
 The stability is characterized by the fidelity for both the same and different initial states.
 We study as initial states the Fock states with definite numbers of atoms in each
 component of the BEC.
 It is found that for some initial times the two Fock states with all the atoms in the same component
 of the BEC are stabler than Fock states with atoms distributed in the two components.
 An experimental scheme is discussed, in which the fidelity can be measured in
 a direct way.

 \end{abstract}

 \maketitle

 \section{Introduction}

 In many research fields, such as  Bose-Einstein condensation (BEC)
 and quantum information processing,
 stable and coherent manipulation of quantum states is of crucial importance \cite{lukin}.
 In fact, the instability issue of BEC  in dilute gases \cite{bec} has been constantly addressed for
 its crucial role in the control, manipulation, and even future
 application of this newly formed matter, including dynamical
 instability \cite{dyinst}, Landau or superfluid instability \cite{Landinst},
 modulation instability \cite{modins,BKPV07}, and quantum fluctuation instability \cite{smerzianglin}.
 It is found  that instability  may break the
 coherence among the atoms and lead to collapse of BEC \cite{liukick}.

 Recently, the stability of the quantum motion of BEC systems under small
 perturbation, which is measured by the so-called quantum Loschmidt echo or
 fidelity, has been studied \cite{pra05-bec,MH08}.
 Here, the fidelity is defined as the overlap of two states obtained by evolving the same initial state
 under two slightly different Hamiltonians \cite{Peres84,nc-book,gc-book}.
 Explicitly, it is  $M(t)=|m(t)|^2$, where
 $m(t)$ is the fidelity amplitude for an initial state $|\Phi_0\ra $, defined as
 \be m(t) = \la \Phi_0|{\rm exp}(i Ht/ \hbar ) {\rm exp}(-i H_0t / \hbar) |\Phi_0 \ra .
 \label{mt} \ee
 Here $H_0$ and $H=H_0 + \epsilon V$ are the unperturbed and perturbed Hamiltonians, respectively,
 with $\epsilon$ a small quantity and $V$ a generic perturbing potential.

 Fidelity decay has been well studied in quantum systems whose classical counterparts
 have chaotic motion
 \cite{JP01,JSB01,BC02,CT02,PZ02,fid-wg,VH03,STB03,Vanicek04,GPSZ06}.
 Related to the perturbation strength, previous investigations show
 the existence of at least three regimes of fidelity decay.
 (i) In the perturbative regime with sufficiently weak perturbation,
 in which the typical transition matrix element is smaller than the
 mean level spacing, the fidelity has a Gaussian decay \cite{Peres84,JSB01,CT02}.
 (ii) Above the perturbative regime, the fidelity has an exponential decay with a rate
 proportional to $\epsilon^2$, usually called the Fermi-golden-rule (FGR) decay of fidelity
 \cite{JP01,JSB01,BC02,CT02,PZ02,fid-wg}.
 (iii) Above the FGR regime is the Lyapunov regime in which $M(t)$ usually has an
 approximate exponential decay with a perturbation-independent rate;
 the decay rate of the fidelity is given by the Lyapunov exponent of the underlying classical dynamics,
 when the classical counterpart of the quantum system has a homogeneous phase space
 \cite{JP01,JSB01,BC02,fid-wg,VH03,STB03}.

 For quantum systems whose classical counterparts have regular motion,
 many investigations in the decaying behavior of fidelity have also been carried out (see, e.g.,
 Ref.~\cite{PZ02,JAB03,PZ03,SL03,Vanicek04,WH05,Comb05,HBSSR05,pra05-bec,GPSZ06,WB06,pre07}),
 however, the situation is still not as clear as in the case of quantum chaotic systems.
 The point is that fidelity decay in quantum regular systems exhibits notable initial-state dependence.
 (In quantum chaotic systems, the main feature of fidelity decay is initial-state independent
 beyond a short initial time.)
 The most thoroughly studied initial states are narrow Gaussian wave packets (coherent states),
 for which the semiclassical theory and numerical simulations show that
 the fidelity has, roughly speaking, a Gaussian decay followed by a long-time power law decay
 \cite{PZ02,WH05,pre07}.
 Some other types of initial states of practical interest
 have also been studied numerically, e.g., the fidelity of an initial maximally
 entangled (N-GHZ) state is shown to have an interesting oscillating behavior \cite{pra05-bec}.

 The above considerations motivate our interest in the stability of a two-component BEC \cite{cornell},
 which is exposed to a pulsed laser field coupling two internal states of the atoms
 in the BEC.
 This BEC system possesses a classical counterpart,
 which has chaotic or regular motion depending on
 both the strength of the coupling field and that of the interaction among the atoms \cite{pra05-bec}.
 In Ref.~\cite{pra05-bec},
 the fidelity of initial coherent states in this BEC system has been studied
 and found in agreement with previous analytical predictions
 in both cases with chaotic and regular motion in the classical limit.
 We remark that, more recently, the fidelity approach has also been employed in the
 study of the stability of another BEC system, which starts from the ground state \cite{MH08}.

 In this paper, we study the stability of the quantum motion of the same BEC system as in
 Ref.~\cite{pra05-bec}.
 However, here we are interested in the stability of initial Fock states,
 which have definite numbers of atoms occupying each of the two components of the BEC,
 and in the parameter regime in which the classical counterpart of the BEC system has regular motion.
 In particular, we are interested in whether some Fock states are more stable than others.
 Knowledge about the stability properties of this type of initial states may be useful for
 potential application of BEC in fields such as quantum information \cite{youli}.
 As in Ref.~\cite{pra05-bec}, we still consider the stability issue caused by small
 perturbation due to imperfect
 control of the coupling field 
 and use fidelity to characterize the stability, with the difference between
 $H$ and $H_0$ in Eq.~(\ref{mt}) given by a small change in the strength of the coupling field.

 Presently, there is no analytical prediction for fidelity decay of initial Fock states,
 therefore, our investigation is mainly based on numerical simulations.
 Interestingly, our numerical results show that, for some initial times, initial Fock states with all the
 atoms in the same component of the BEC are more stable than other Fock states.
 To see whether or not this property is due to the specific measure given in Eq.~(\ref{mt}),
 we have also studied the behavior of a more general form of the fidelity,
 which is the overlap of the time evolution of two different initial Fock states under
 two slightly different  Hamiltonians.
 Our numerical simulations for this more general fidelity give consistent results.

 The paper is organized as follows:
 In the second section, we discuss briefly the two-component BEC model.
 Section III is devoted to a study of fidelity for the same initial Fock states.
 In Sec.~IV, we introduce and study numerically the more general fidelity mentioned above
 for two different initial states.
 Conclusions are given in Sec.~V.
 An experimental scheme for measuring fidelity of the quantum motion of a
 two-component BEC is discussed in the Appendix.

\section{Physical Model}

 We consider the same BEC system as in Ref.~\cite{pra05-bec},
 specifically, cooled $^{87}$Rb atoms
 with two different hyperfine states $F=1,m_F=-1$ and $F=2,m_F=+1$.
 The total number of the atoms in the BEC is $N$.
 A near resonant, pulsed radiation laser field is  used to couple the two internal states.
 Within the standard rotating-wave approximation, the Hamiltonian
 describing the transition between the two internal states reads
 \bey \nonumber
 \hat{H}=\frac{\mu}{2} (\hat{a}_1^{\dagger }\hat{a}_1-\hat{a}_2^{\dagger }\hat{a}%
 _2)+\frac{g}{4}(\hat{a}_1^{\dagger }\hat{a}_1-\hat{a}_2^{\dagger }\hat{a}_2)^2
 \\ +\frac{K}{2}
 (\hat{a}_1^{\dagger }\hat{a}_2+\hat{a}_2^{\dagger }\hat{a}_1) \sum_n\delta (t-nT),
 \eey
 where $K$ is the coupling strength proportional to the laser field.
 Here we suppose that the laser field used to couple the two states is turned
 on only at certain times with a period $T$. The operators  $\hat{a}_1,\, \hat{a}_1^{\dagger
 },\,\hat{a}_2$, and $\hat{a}_2^{\dagger }$ are boson annihilation and
 creation operators for the two components, respectively. The parameters are $K= \hbar
 \Omega _R,\, g=\frac{2\pi \hbar^2}{m}\eta (2a_{12}-a_{11}-a_{22}),\,\mu =-\delta +(4N\pi
 \hbar^2/m)\eta (a_{11}-a_{22}).$ Here, $\Omega _R$ is the Rabi
 frequency; $a_{ij}$ is the $s$-wave scattering amplitude; $\delta $ is the detuning
 of lasers from resonance, very small and negligible in our case;
 $m$ is the mass of atom; $\eta $ is a constant of order one independent of
 the hyperfine index, relating to an integral of equilibrium
 condensate wave function \cite{leggett}.

\begin{figure}
\includegraphics[width=\columnwidth]{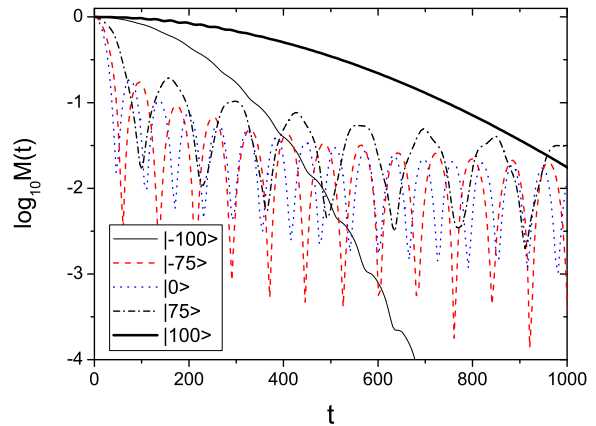}  \vspace{-0.2cm}
 \caption{ Decay of fidelity for $K=1$, $g_c=0.2$, $L=100$, and $\sigma = 0.1$.
 The initial states are Fock states $|l\ra $ of $l=-100$ (thin solid curve), -75 (dashed curve),
 0 (dotted curve), 75 (dashed-dotted curve), and 100 (thick solid curve), respectively.
 The initial decay of the fidelity of the two  states $|L\ra $ and $|-L\ra $ is much slower than
 that of the other three Fock states.   }
 \label{s0p1}
 \end{figure}

 The above Hamiltonian can be written in terms of the SU(2)
 generators \cite{anglin},
 \be \hat{L}_x=\frac{\hat{a}_1^{\dagger
 }\hat{a}_2+\hat{a}_2^{\dagger }\hat{a}_1}2, \hat{L}
 _y=\frac{\hat{a}_1^{\dagger }\hat{a}_2-\hat{a}_2^{\dagger
 }\hat{a}_1}{2i}, \hat{L}_z=\frac{\hat{a}_1^{\dagger
 }\hat{a}_1-\hat{a}_2^{\dagger }\hat{a}_2} 2,\ee
 which gives
 \be  \hat{H}=\mu \hat{L}_z+g\hat{L}_z^2+K\delta _T(t)\hat{L}_x . \label{H-L} \ee
 The Floquet operator describing the quantum evolution in one period is \cite{pra05-bec,PZ02}
 \begin{equation}
 \hat{U}=\exp [-i(\mu \hat{L}_z+g\hat{L}_z^2)T]\exp (-iK\hat{L}_x),
 \end{equation}
 where the Planck constant is set unit unless otherwise stated.
 Since the overall scaling of the Hamiltonian does not influence dynamical properties of the
 system, we will set $\mu $ unit.

 The Hilbert space for the system is spanned by the eigenstates of $\hat{L}_z$,
 denoted by $|l\ra $ with $l=-L, -L+1, \ldots , L$, where $L=N/2$.
 These states $|l\ra $ are the Fock states.
 Using $N_1$ and $N_2$ to denote the numbers of the atoms in the two components, respectively,
 with $N=N_1+N_2$, we have $l =(N_1-N_2)/2 $.
 Hence, $|l\ra $ is the state with $N_1=(L+l)$ atoms in the first component and
 with $N_2=(L-l)$ atoms in the second component.
 In particular, the two states with all the atoms in one of the two components
 are $|-L\ra $ and $|L\ra $.
 The SU(2) representation of the system discussed above is quite convenient
 for the study of properties of the Fock states $|l\ra $.

\begin{figure}
\includegraphics[width=\columnwidth]{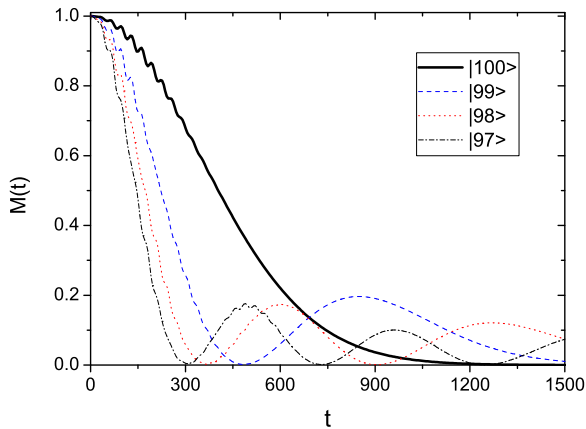}  \vspace{-0.2cm}
 \caption{ Decay of fidelity for $K=1, g_c=0.2, L=100$, and $ \sigma = 0.1$.
 The initial states are Fock states $|l\ra $ of $l=100,99,98$
 and 97, respectively.    }
 \label{97-100}
 \end{figure}

 Note that for some specific choice of the parameters
 the system degenerates to the quantum kicked top model \cite{kicktop}.
 As in the kicked top model, an effective Planck constant can be introduced,
 $\hbar_{\rm eff} = 1/L$, which will be written as $\hbar $ in what follows for brevity.
 The system has a classical counterpart in the limit $N \to \infty $.

 \section{Fidelity decay for initial Fock states}
 \label{sect-fid}

 As mentioned in the Introduction, we consider a small perturbation due to imperfect
 control of the coupling field.
 Specifically, for an unperturbed Hamiltonian $H_0$ with the form given on the right-hand
 side of Eq.~(\ref{H-L}), the perturbed Hamiltonian ($H_0+\epsilon V$)
 is given by the change $K\to K+\epsilon $.
 Denoting the one-period evolution operators corresponding to the unperturbed and the perturbed
 Hamiltonians by $\hat U$ and $ \hat{U}_\epsilon $, respectively,
 the fidelity amplitude $m(t)$ is now written as
 \begin{equation}
 m(t=nT)=\langle \Phi _0| \left ( \hat{U}_\epsilon ^{\dagger } \right )^n \circ
 \left ( \hat{U} \right )^n |\Phi _0\rangle ,
 \label{3} \end{equation}
 with $|\Phi _0\rangle $ indicating the initial state.
 Fast decay of the fidelity means rapid lose of information during the quantum evolution
 in the presence of the perturbation.
 For small $\epsilon $, it is usually convenient to use $\sigma =\epsilon / \hbar $
 as a measure for the strength of quantum perturbation.

\begin{figure}
\includegraphics[width=\columnwidth]{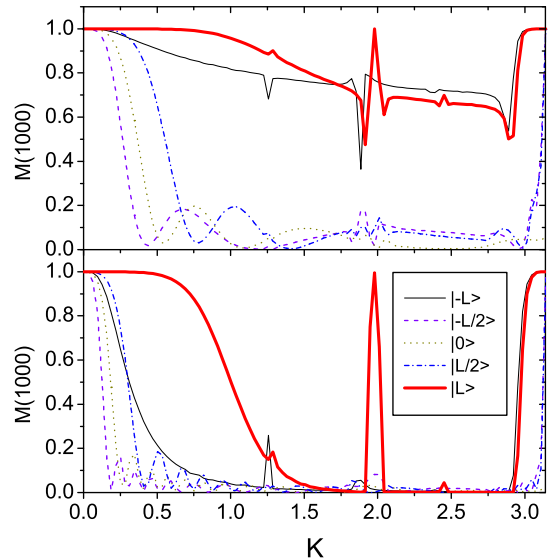}  \vspace{-0.2cm}
 \caption{ Values of $M(t)$ for $t=1000$, as a function of $K$.
 The five curves correspond to five initial Fock states: the thin solid curve for $|-L\ra $,
 dashed curve for $|-L/2\ra $, dotted curve for $|0\ra $, dashed-dotted curve for $|L/2\ra$,
 and thick sold curve for $|L\ra $.
 Parameters are $L=100, g_c=0.2$, and $\sigma = 0.01$ for the upper panel
 and $\sigma =0.04$ for the lower panel.   }
 \label{M1000-gc2}
 \end{figure}

 In this paper, we consider only the parameter regime in which the corresponding
 classical system has regular motion (see Ref.~\cite{pra05-bec} for details of the regime).
 We have carried out numerical investigations in the fidelity decay of initial Fock states
 $|l\ra $ in this parameter regime.
 It is found that the two Fock states with $|l|=L$, i.e., with all the atoms occupying the same
 component of the BEC, behave differently from other Fock states.

 Some examples of our simulations are shown in Fig.~\ref{s0p1},
 with parameters $K=1$, $g_c\equiv gL=0.2$, $L=100$, and $\sigma = 0.1$.
 For some initial times, the fidelity $M(t)$ of the two initial Fock states
 with $|l|=L$ has a decay which is approximately a Gaussian decay and is much slower than
 the fidelity decay of the other three Fock states.
 For long times,  the fidelity of $|L\ra $ and $|-L\ra $
 are smaller than that of the other three Fock states,
 the latter of which oscillates and decays slowly on average.
 Similar results have also been found for other Fock states $|l\ra$ with $|l|$ not close to $L$.
 We have varied the perturbation strength, with $\sigma $ from $0.01$ to $5$,
 and found qualitatively similar results.
 These results show that for not long times,
 the two Fock states $|L\ra $ and $|-L\ra $ are more stable than other Fock states
 with $|l|$ not close to $L$.

 There is also some difference between the fidelity decay of the two initial states
 $|-L>$ and $|L>$, as shown in Fig.~\ref{s0p1}, that is, the fidelity of $|L\ra $ decays more slowly
 than that of $|-L\ra $.
 This can be understood from the form of the Hamiltonian in Eq.~(\ref{H-L}).
 Indeed, for the state $|L\ra $, the first two terms on the right-hand side of Eq.~(\ref{H-L})
 give $(L + g_c L)$, while for $|-L\ra $ they give $(g_cL-L)$.
 Since $L + g_c L >|g_cL-L|$,
 the state $|L\ra $ is less perturbed than the state $|-L\ra $ for the same change of
 the parameter $K$.

 We have also studied fidelity decay of Fock states $|l\ra $ with $|l|$ close to $L$.
 Figure \ref{97-100} gives the cases of $l=100,99,98$, and 97,
 which shows that the rate of the initial decay of fidelity increases with decreasing $l$,
 and the oscillation of $M(t)$ appears even at $l=99$.
 For $l$ close to $-L$, the situation is similar.

\begin{figure}
\includegraphics[width=\columnwidth]{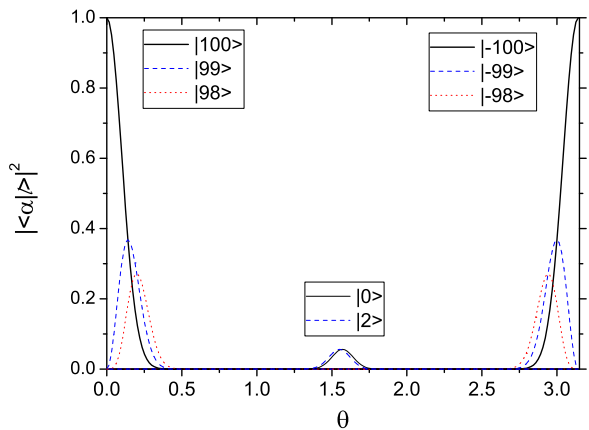}  \vspace{-0.2cm}
 \caption{
 Variation of $|\la \alpha |l\ra |^2$ with $\theta$ for $L=100$,
 where $|\alpha \ra $ is the coherent state in Eq.~(\ref{alpha}).
 For the three curves on the left, $l=100$ (solid curve), 99 (dashed curve)
 and 98 (dotted curve), respectively.
 For the three on the right, $l=-100,-99$ and -98, respectively.
 For the two in the middle, $l=0$ and 2, respectively.
 } \label{expan-cohe}
 \end{figure}

 In Fig.~\ref{M1000-gc2}, we show fidelity decay at different values of $K$ corresponding to
 the same fixed time $t$.
 For $\sigma =0.01$ shown in the upper panel,
 at $t=1000$ the fidelity of the two Fock states $|L\ra $ and $|-L\ra $
 is much higher than that of the other three Fock states for $K$ between 0.5 and 3.
 With increasing the perturbation strength $\sigma$, the fidelity at this time becomes smaller.
 For $\sigma =0.04$ shown in the lower panel, $M(1000)$ of $|-L\ra $ is already small,
 while $M(1000)$ of $|L\ra $ is still high within some windows of $K$.
 This is because the fidelity of $|-L\ra $ decays faster than that of $|L\ra $
 (see Fig.~\ref{s0p1} and discussions in a previous paragraph).

 In quantum regular systems, the fidelity of initial
 coherent states is known to have initial Gaussian decay \cite{PZ02}.
 Hence, it is natural to check the relationship between the Fock states and the coherent states
 generated by generators of the group SU(2).
 A SU(2) coherent state  $|\alpha\ra $ centered at a point in the sphere with polar angle $\theta$
 and azimuthal angle $\phi$ is given by \cite{Perelomov,ZFG90}
 \be \label{alpha} |\alpha \rangle  \equiv e^{\alpha^{*}\hat L_{+}-\alpha \hat L_{-}}|-L\rangle ,
 \ \  \text{with} \ \alpha =\frac{\pi -\theta }2e^{-\mathrm{i}\phi }. \ee
 The state $|-L\ra $ is a coherent state with $\alpha =0$.
 To see the relation of other Fock states to coherent states,
 one can expand $|\alpha \ra $ in the Fock states
 \be |\alpha \ra = \sum_{l=-L}^{L} \frac{(z^*)^{l+L}}{(1+zz^*)^L} \left [
 \frac{(2L)!}{(L+l)! (L-l)!} \right ]^{1/2} |l\ra  , \ee
 where $ z=- e^{-i\phi } \cot (\theta /2)$.
 This gives
 \be |\la \alpha |l\ra |^2 = \frac{|z|^{2(l+L)}}{(1+|z|^2)^{2L}} \left [
 \frac{(2L)!}{(L+l)! (L-l)!} \right ] . \ee
 Considering the limit $|z|\to \infty$, it is easy to check that $|L\ra $ is also a coherent state.
 Other Fock states are not coherent states; some examples are shown in Fig.~\ref{expan-cohe}.

 One may expect that, since the two Fock states $|L\ra $ and $|-L\ra $ are coherent states,
 the initial slow Gaussian decay of their fidelity, as shown in Fig.~\ref{s0p1},
 might be explained quantitatively by making use to results given in Ref.~\cite{PZ02}.
 However, detailed analysis show that the situation here is more complex.
 Indeed, in the derivation of Gaussian decay given in Ref.~\cite{PZ02},
 a property of coherent states is made use of,
 namely, a Gaussian form of the expansion of a coherent state in the basis of Fock states.
 This property is possessed by most of the coherent states,
 but the two states $|L\ra $ and $|-L\ra $ are themselves Fock states.
 If one tries to use a $\delta $-function for the expansion,
 going along the line of Ref.~\cite{PZ02}, it is found that their fidelity has a decay rate which is
 equal to zero.
 Hence, this approach can not explain quantitatively the initial Gaussian decay of the
 fidelity of the two states $|L\ra $ and $|-L\ra $.
 But, it indeed gives a qualitative explanation to the fact that the decay is slow.

 \section{Fidelity for different initial states}

 In practical situations, experimentally prepared initial states are usually not exactly
 the same as the expected ones.
 Hence, in addition to perturbation, one should also consider small change in
 the initial state in the study of fidelity,
 i.e., considering fidelity of a more general form than that in given Eq.~(\ref{mt}).
 In this section, we consider such a more general fidelity
 and use it in the study of the stability of Fock states.

\begin{figure}  \includegraphics[width=\columnwidth]{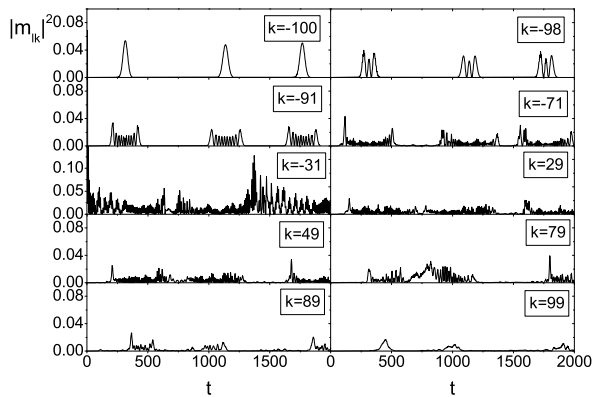}  \vspace{-0.2cm}
 \caption{ Variation of $|m_{lk}(t)|^2$ with t for $l=-31$.
 It has relatively regular peaks when $|k|$ is close to $L$.
 Parameters are $L=100, K=2, g_c=0.17$, for which the classical
 counterpart has regular motion, and $\sigma = 0.5$.}
 \label{ML70K} \end{figure}

 \subsection{Fidelity for different initial Fock states}

 Suppose the expected initial state is a Fock state $|k\ra $, while what is really prepared is the state
 \be |\Psi_0 \ra  = \sum_{l } C_{l } | l \ra  \ee
 with $|C_k|$ close to but smaller than one.
 The expected state at time $t$ is the time evolution of $|k\ra $ under $H_0$,
 while the real state is the time evolution of $|\Psi_0\ra $ under $H$.
 Hence, one should consider the following more general form of the fidelity amplitude
 (see Ref.~\cite{nc-book}),
 \bey \nonumber f(t) = \la \Psi_0 | e^{iHt} e^{-iH_0t} |k\ra = \sum_l  C_l^* m_{lk}(t), \eey
 where
 \be m_{lk}(t) = \la l| e^{iHt } e^{-iH_0t} |k \ra . \ee

 The quantity $m_{lk}(t)$ can be regarded as a generalized echo, from an initial state $|k\ra $
 to a final state $|l\ra $.
 Its absolute value square, $ M_{lk}(t)\equiv |m_{lk}(t)|^2$ gives the probability for the
 final state to be found in $|l\ra $,
 if the initial state is $|k\ra $ and the dynamics is governed
 by  $H_0$ for the first time interval $t$ and by $(-H)$ for the second interval with the same length.
 For $k=l$, $m_{lk}(t)$ is just the fidelity amplitude in Eq.~(\ref{mt}).
 In this section, we are more interested in the case of $k \ne l$,
 for which $|m_{lk}(0)|^2=0$.

 For a quantum regular system $H_0$, $ M_{lk}(t)$ of $k \ne l$ may be considerably large
 within some time windows, due to the peculiarity of integrability.
 [If $H_0$ is a quantum chaotic system, $M_{lk}(t)$ usually approaches its saturation value soon
 and fluctuates around the saturation value.]

 We would like to mention that the quantity $m_{lk}(t)$ also appears,
 when one considers the difference between the expectation values of an arbitrary observable $A$
 for the same initial state under two slightly different Hamiltonians.
 Indeed, consider an initial state $|k\ra $ and the expectation values of $A$ in the two systems
 \bey A_{kk}^{H_0} (t) \equiv \la k| e^{iH_0 t} A e^{-iH_0 t} |k \ra , \label{AH0t}
 \\ A_{kk}^{H} (t) \equiv \la k| e^{iH t} A e^{-iH t} |k \ra . \label{AHt} \eey
 Inserting the identity operator
 $ \sum_l e^{-iH_0t} |l\ra \la l | e^{iH_0t} $
 in Eq.~(\ref{AHt}) before and after the operator $A$, we obtain
 \be  A_{kk}^{H}(t) - A_{kk}^{H_0}(t)  = {\sum_{ll'}}' m_{kl}(t) m_{kl'}^*(t) A_{ll'}^{H_0} (t), \ee
 where the prime over the sum implies that $l=k$ and $l'=k$ can not hold at the same time and
 \be  A_{ll'}^{H_0} (t)  \equiv \la l| e^{iH_0 t} A e^{-iH_0 t} |l' \ra  \ee
 is a quantity given in the system $H_0$.


\begin{figure}[!t]
\includegraphics[width=\columnwidth]{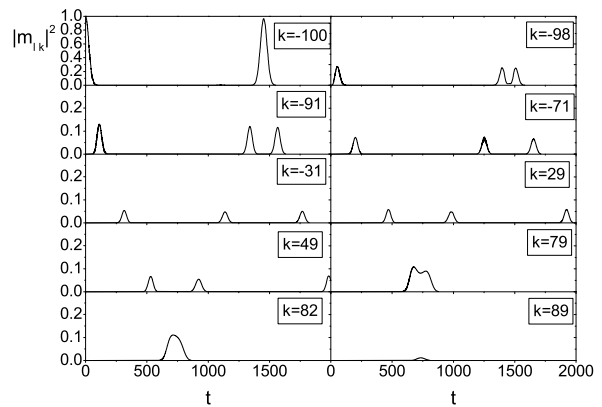}  \vspace{-0.2cm}
 \caption{ Same as in Fig.\ref{ML70K} for $l=-L=-100$. }
 \label{ML1K} \end{figure}


 \subsection{Numerical investigation }

 Now we study the behavior of the quantity $M_{lk}(t)$
 in the two-component BEC system discussed above.
 As shown in Sec.~\ref{sect-fid}, for some initial times
 the two Fock states $|L\ra $ and $|-L\ra$
 are more stable than other Fock states $|l\ra $ with $|l|$ not close to $L$,
 in the sense that the former's fidelity is higher than the latter's.
 In this section, we show that our numerical investigations in the quantity $M_{lk}(t)$
 give consistent results.
 Specifically, $M_{lk}(t)$ with either $|l|$ or $|k|$ close to $L$ behave
 more regularly than those with both $|l|$ and $|k|$ far from $L$.

 Let us first consider $M_{lk}(t)$ with $|l|$ far from $L$.
 An example is given in Fig.~\ref{ML70K},
 which shows variation of $M_{lk}(t)$ with time $t$ for $L=100$,  $l=-31$,
 and $k$ from -100 to 99.
 It shows that for $k$ close to $-L$, $M_{lk}(t)$ has quite regular peaks,
 and with increasing $k$ from $-L$ to $0$, the peaks become more and more irregular.
 Similarly, for $k$ decreasing from $L$ to 0, the peaks of $M_{lk}(t)$ also becomes
 more and more irregular.

\begin{figure}
\includegraphics[width=\columnwidth]{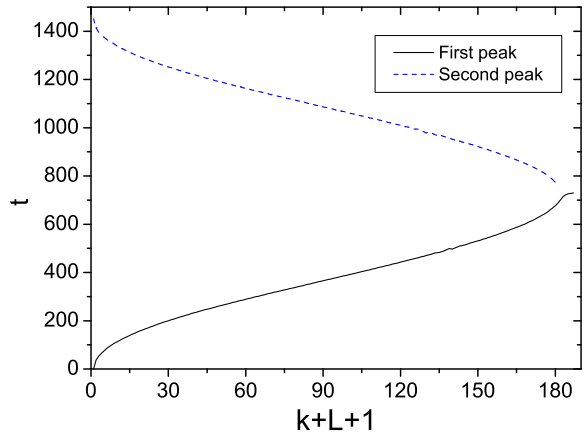}  \vspace{-0.2cm}
 \caption{Variation with $k$ for the times corresponding to the centers of the first and second
 peaks of $|m_{lk}(t)|^2$ shown in Fig.~\ref{ML1K}. }
 \label{max-tk} \end{figure}

 On the other hand, for $l=-L=-100$, $M_{lk}(t)$ behaves regularly for all the values
 of $k$, as shown in Fig.~\ref{ML1K}.
 Here, the basic feature is that for each value of $k$, $M_{lk}(t)$ is considerably large
 only within some time intervals.
 Specifically, for $k=-100$, $M_{lk}(t)=1$ at $t=0$ as a trivial result of $k=l$,
 then, it decays and remains small until $t\simeq 1370$ after which there is a revival,
 forming a second peak centered at $t\simeq 1450$.
 The second peak of $M_{lk}(t)$ of $k=-100$ splits into two peaks when $k=-98$,
 hence, $M_{lk}(t)$ has three peaks for $k=-98$.
 Interestingly, with further increasing $k$, the first peak of $M_{lk}(t)$
 moves to the right while the second peak moves to the left.
 The two peaks meet at a value of $k$ a little smaller than 79.
 In Fig.~\ref{max-tk}, we show variation of the times corresponding to
 the centers of the first and second peaks with increasing $k$.

\begin{figure}
\includegraphics[width=\columnwidth]{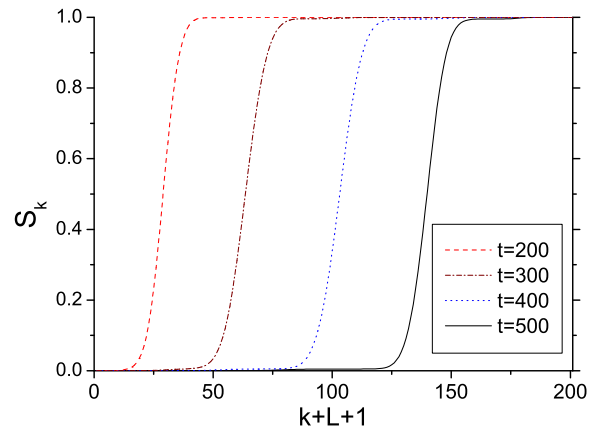}  \vspace{-0.2cm}
 \caption{Values of $S_k = \sum_{k'\le k} M_{lk'}(t)$ for $l=-L$.
 Parameters are the same as in Fig.\ref{ML70K}. It shows that at each of the times, $M_{lk}(t)$
 is concentrated in a small region of $k$.   }
 \label{Int-k-L1} \end{figure}

 The structure of the peaks shown in Fig.~\ref{ML1K} suggests that at each fixed time $t$,
 $|M_{lk}(t)|$ of $l=-L$ may be concentrated in a relatively small region of $k$.
 Indeed, this is confirmed by our further numerical simulations.
 We have calculated the quantity
 \be S_k(l,t) = \sum_{k'\le k} M_{lk'}(t), \label{Sk} \ee
 which gives the total probability for $k' \le k$.
 Examples for some fixed times are given in Fig.~\ref{Int-k-L1},
 which shows that $|M_{lk}(t)|$ of $l=-L$ is indeed concentrated in a relatively narrow region of $k$
 for each time.

 \section{Conclusions}

 By numerical simulations, we have studied quantum Loschmidt echo or fidelity decay
 of initial Fock states in a two-component
 BEC system, whose classical counterpart has regular motion.
 Our results show that, for some initial times, initial Fock states with all the atoms in one component of
 the BEC are more stable than Fock states with atoms distributed in the two components.
 This implies that one-component BEC might be more stable than two-component BEC.

 We have further investigated this issue by considering a more general form of the fidelity, i.e.,
 fidelity for different initial states.
 Numerical computations of the general fidelity show consistent results,
 namely, initial Fock states with all the atoms in the
 same component behave more regularly than other Fock states.

\acknowledgments

One of the authors (W.W.) is partially supported by National Natural Science Foundation of China Grants
No.~10775123 and No.~10275011.
Two of the authors (W.W. and B.L.) are partially supported
by an Academic Research Fund of NUS.
One of the authors (J.L.) is supported by National Natural Science Foundation of China
(Grant No.~10725521), the National Fundamental Research Programme of China
under Grants No.~2006CB921400, No.~2007CB814800.

 \appendix
 \section{An experimental scheme for measuring fidelity}

 In this appendix, we discuss in detail an experimental scheme for measuring fidelity decay
 in a BEC system, which  is briefly sketched in the last section of Ref.~\cite{pra05-bec},
 and give an explicit expression of the fidelity in terms of measurable quantities.

 Experimental schemes for measuring fidelity have been discussed by several groups
 and basically three types of schemes have been proposed \cite{GPSZ06}.
 In the first type of scheme, a quantum system is considered, which is composed of two subsystems
 (or has two degrees of freedom) \cite{GCZ97,HBSSR05,WB06,PD07}.
 The system is assumed to have a time evolution such that the
 fidelity of the first subsystem is given by the reduced density matrix of the second subsystem
 \cite{GPSS04,QSLZS06}.
 Then, measuring properties of the second subsystem which can be small,
 fidelity of the first subsystem which may be large can be obtained.
 This scheme is adopted in the experiments in Ref.~\cite{AKD03,REPNL05}.
 To be specific, one may consider a Hilbert space which
 is the direct product of a two-dimensional subspace with
 basis states $|1\ra $ and $|2\ra $ and a second subspace for state vectors $|\psi \ra $.
 The Hamiltonian has the form $H = H_1 |1\ra \la 1| + H_2 |2\ra \la 2|$, where
 $H_1$ and $H_2$ act in the second subspace only.
 For an initial state $|\Psi (0)\ra = (|1\ra + |2\ra )|\psi_0 \ra  / \sqrt 2$,
 the state at time $t$ is
 $ |\Psi (t)\ra = \frac{1}{\sqrt 2}\left ( e^{-iH_1t} |\psi_0 \ra |1\ra  +
 e^{-iH_2t} |\psi_0 \ra |2\ra \right )$ .
 Then, the fidelity amplitude $\la \psi_0| e^{iH_2t} e^{-iH_1t} |\psi_0 \ra$ in the second
 subspace is related to an off-diagonal
 element of the reduced density matrix in the first subspace.

 In the second type of scheme, a special kind of system is considered,
 for which the time evolution can be such controlled
 that the system evolves under $H_0$ for a time period $t$, then,
 evolves under  $(-H_{0}-\epsilon V)$ for a second period $t$.
 The fidelity is then just the survival probability, i.e., the probability for the final state to be
 found in the initial state.
 In the third type, classical waves are employed, which evolves according to a dynamical
 law mathematically equivalent to Schr\"{o}dinger equation \cite{SSGS05}.

 Now we discuss the scheme briefly mentioned in Ref.~\cite{pra05-bec}.
 We propose to use a setup similar to that used in Ref.~\cite{ketterle}.
 Consider a BEC (e.g.~$^{87} Rb$) which is optically cooled and trapped, then,
 transferred into a double-well potential.
 The  double-well potential can be created by
 deforming a single-well optical trap into a double-well
 potential with linearly increasing the frequency difference between the rf signals \cite{ketterle}.
 Near-resonant coupling fields are applied to the BEC in the two wells with slight
 difference in strength.
 Finally, simultaneously switching off all the external fields
 and letting the BEC expand freely, interference pattern of the BEC can be observed.
 The wells should be deep, such that the total density
 remains approximately a constant, and the atom numbers in the two wells are required nearly equal.

 At the initial time $t_0$, suppose the state of the system is a product state
 $ |\Psi (t_0)\ra = |\phi (t_0)\ra |\psi (t_0)\ra |\Phi_p (t_0) \ra ,$
 where $|\phi (t_0)\ra$ is the internal state of the atoms,
 e.g., with all the atoms in the same hyperfine internal state,
 $|\psi (t_0)\ra $ describes the motion of the center-of-mass degrees of freedom of the atoms,
 and $|\Phi_p (t_0) \ra $ represents the field forming the optical trap.
 We assume that the field of the optical trap is not entangled with the BEC in the experimental
 process and shall omit the term $|\Phi_p (t) \ra $.

 From time $t_0$ to $t_1$, the potential of the optical trap is deformed into a double-well potential.
 If the internal state of the atoms is not influenced in this process,
 at $t=t_1$, one has
 $ |\Psi (t_1)\ra = |\phi (t_1)\ra \left [ |\psi_1 (\bR_1,t_1)\ra +
 |\psi_2 (\bR_2,t_1)\ra \right ] ,$
 where $\bR_1$ and $\bR_2$ indicate spatial locations of the two wells, respectively.

 From time $t_1$ to $t_2$,
 near-resonant coupling fields can be applied to the
 condensates  to couple the two hyperfine states.
 The coupling fields have a slight difference in strength in the two wells.
 We assume that the near-resonant coupling fields can be treated as classical fields
 and do not induce tunnelling between the two wells.
 The internal states of the condensate in the two wells will then evolve differently.
 Thus, for $t\in (t_1,t_2)$,
 \bey  |\Psi (t)\ra =   |\phi_1 (t)\ra  |\psi_1 (\bR_1,t)\ra
  + |\phi_2 (t)\ra  |\psi_2 (\bR_2,t)\ra . \label{pp1} \eey
 In the case that the internal degrees of freedom are not coupled to the center-of-mass degrees of freedom,
 $|\phi_j (t)\ra$ has unitary time evolution
 \be |\phi_j (t)\ra = U_j(t_1,t)|\phi (t_1)\ra , \label{phi-U} \ee
 with $ j=1,2$ indicating the two wells.
 The internal state can be expanded in the Fock states $|l\ra $,
 \be |\phi_j (t)\ra = \sum_l d_l^{(j)}(t) |l\ra . \label{phi-l} \ee

 At $t=t_2$, one can simultaneously switch off all the external fields
 and let the two clouds of BEC expand freely.
 For $t> t_2$,  Eqs.~(\ref{pp1}) and (\ref{phi-l}) are still valid.
 Substituting Eq.~(\ref{phi-l}) into Eq.~(\ref{pp1}), one has
 \bey |\Psi (t)\ra = \sum_l \left [ d_l^{(1)}(t) |\psi_1 (\bR_1,t)\ra
 + d_l^{(2)}(t)  |\psi_2 (\bR_2,t)\ra \right ] |l\ra .\ \  \eey
 Suppose the single particle states for $ |\psi_1 (\bR_1,t)\ra$ and $ |\psi_2 (\bR_2,t)\ra$
 are $\chi_1 (\bx ,t)$ and $\chi_2 (\bx ,t)$, respectively.
 Then, the probability of finding a particle at a position $\bx $ is
 \bey  P(\bx ,t)
  = |\chi_1|^2 + |\chi_2|^2 + 2{\rm Re} \left [ \ww f(t) \chi_1 \chi_2^*
 \right ] , \label{P-f} \eey
 where $ \ww f(t) \equiv  \sum_l d_l^{(1)}(t) {d_l^{(2)}}^*(t) $.

 Making use of Eqs.~(\ref{phi-U}) and (\ref{phi-l}), it is seen that
 \be \ww f(t)  = \la \phi(t_1)|U_2^{\dag}(t_1,t) U_1(t_1,t)|\phi(t_1)\ra , \ee
 which is a fidelity amplitude.
 Since there is no coupling field beyond $t_2$, $U_1(t_2,t)= U_2(t_2,t)$
 and, as a result, $\ww f(t)=\ww f(t_2)$ for $t>t_2$.
 Then, Eq.~(\ref{P-f}) can be written as
 \be 2|\ww f(t_2)| {\rm Re} \left [ e^{i\theta_f(t_2)} \chi_1 \chi_2^* \right ]
 = P(\bx ,t) - |\chi_1|^2 - |\chi_2|^2, \ee
 where $\theta_f(t_2)$ is the phase of $\ww f(t_2)$.
 Therefore, the value of $|\ww f(t_2)|$ can be obtained by measuring the interference
 pattern of the two expanding clouds of BEC.

\end{document}